\documentstyle[preprint,prl,aps]{revtex}

\begin{document}
\draft
\preprint{{\vbox{\hbox {UR-1538} \hbox{rev. Feb 1999}}}}

\title{Neutrino Oscillations and Flavor Structure of 
Supersymmetry without R-Parity}
\author{\bf Otto C. W. Kong
}
\address{Department of Physics and Astronomy,\\
University of Rochester, Rochester NY 14627-0171}
\maketitle

\begin{abstract}
Focusing on a simple three neutrino oscillation scenario motivated by the 
recent Super-Kamiokande result, we discuss both tree-level and 1-loop
contributions to neutrino masses in supersymmetry without R-parity,
aiming at discerning the flavor structure issues of the theory.
The single-VEV parametrization framework, which allows the first
consistent treatment of the bilinear and trilinear R-parity violating terms, 
and hence the tree and 1-loop contributions, without any assumption
on the nature of R-parity violation. Though the very small neutrino 
masses implies stringent suppressions of the relevant R-parity violating 
couplings, we show that there is still room for understanding the 
suppressions as a simple consequence of the general flavor hierarchy. 
\end{abstract}

\newpage

The recent zenith angle dependence measurement by the Super-Kamiokande 
(SuperK) experiment\cite{SuperK} has drawn much attention to 
neutrino oscillations, particularly the scenario of three neutrino 
oscillations explaining the SuperK result and the solar neutrino 
deficit\cite{nu}. The scenario is described by
\begin{eqnarray*}
\Delta\!m^2_{\rm atm} &\simeq& (0.5-6)\times 10^{-3}\, \mbox{eV}^2 \\
\sin^2\!2\theta_{\rm atm}   &\simeq& (0.82-1) \\
  \Delta\!m^2_{\rm sol} &\simeq& (4-10)\times 10^{-6}\, \mbox{eV}^2 \\
\sin^2\!2\theta_{\rm sol}   &\simeq& (0.12-1.2)\times 10^{-2}
\end{eqnarray*}
with $\nu_\mu - \nu_\tau$ to be responsible for the Super-K atmospheric 
result and MSW-oscillation of $\nu_e$ for the solar neutrino problem. 
The most natural setting then would be for the two neutrino mass
eigenvalues of the $\nu_\mu - \nu_\tau$ system to have $m^2 \simeq
  \Delta\!m^2_{\rm sol}$ and $  \Delta\!m^2_{\rm atm}$. We will
concentrate on this particular scenario below.

When supersymmetry (SUSY) is adjointed to the standard model (SM), lepton
number violation couplings naturally give rise to neutrino masses, unless
the couplings are banned, such as by imposing R-parity. The subject is
of much interest on its own\cite{NP,nurr,nuR}. In a recent paper\cite{CKKL},
the generation of exactly the above three neutrino oscillation scenario from 
R-parity violation is analyzed. The analysis has unification scale
assumptions on R-parity violation and makes some use of the flavor
structure of the R-parity violating (RPV) couplings. In our opinion,
flavor structure of the full theory of SUSY without R-parity could be a
tricky problem and deserves more attention.

Most of the RPV couplings have to be small\cite{rpv},
compare with their R-parity conserving counter-parts. However, apart 
from the baryon violating couplings, their smallness may not be 
particularly remarkable compared with that of the standard Yukawa
couplings giving rise to masses of the lighter-two-family fermions.
In fact, the suggestion that the smallness of all the couplings may be
understood from a simple approximate flavor symmetry perspective
had been made\cite{H}. Nevertheless, an explicit analysis of the type
is missing. Actually,  a  framework that allows the explicit 
phenomenological studies of the {\it complete} supersymmetric model without 
R-parity, {\it i.e.} with no {\it a priori} assumption on vanishing of any
of the admissible RPV couplings taken, is presented only in a recent
paper\cite{I}.

The present paper is the first attempt at an explicit 
analysis of the feasibility of an approximate flavor symmetry 
perspective on the complete supersymmetric standard model
without R-parity. This is done here, only in the limited context of fitting
the above described three neutrino oscillation scenario. Besides the
great interest in the latter, our analysis also illustrates some basic issues
in the flavor structure of SUSY without R-parity that would be useful for
further studies on the topic. Naively,  the neutrino oscillation scenario,
with the very small neutrino masses, seems to indicate a very stringent
suppresion in R-parity violation\cite{BFK}. So this limited context,
though relatively simple, may actually represent an extreme case of
small  R-parity violation and hence the difficult end of the spectrum 
for the approximate flavor symmetry perspective. Result of our analysis,
hence, serves as a strong indicator for the more general cases.
We should point out that in a complementary 
perspective, there has also been some horizontal (family) symmetry 
model-building works addressing various aspect of R-parity 
violation\cite{B,nuf}. 

The most general renormalizable superpotential for the supersymmetric
SM without R-parity can be written as
\begin{equation}
W = \varepsilon_{ab}\left[ \mu_{\alpha}  \hat{L}_{\alpha}^a \hat{H}_u^b
+ h_{ik}^u \hat{Q}_i^a   \hat{H}_{u}^b \hat{U}_k^{\scriptscriptstyle C}
+ \lambda_{i\alpha k}^{'} \hat{Q}_i^a \hat{L}_{\alpha}^b 
\hat{D}_k^{\scriptscriptstyle C} + \lambda_{\alpha \beta k}  \hat{L}_{\alpha}^a  
 \hat{L}_{\beta}^b \hat{E}_k^{\scriptscriptstyle C}
\right] + \lambda_{i jk}^{''}  \hat{D}_i^{\scriptscriptstyle C}  
\hat{D}_j^{\scriptscriptstyle C}  \hat{U}_k^{\scriptscriptstyle C}\; ,
\end{equation}
where  $(a,b)$ are $SU(2)$ indices, $(i,j,k)$ are family (flavor) indices,
while $(\alpha,\beta)$ are (extended) flavor indices from $0$ to $3$
with  $\hat{L}_{\alpha}$'s denote the four doublet superfields with $Y=-1/2$.  
$\lambda$ and $\lambda^{''}$ are antisymmetric in the first two indices as
required by  $SU(2)$ and  $SU(3)$ product rules respectively,
though  only the former is shown explicitly here, 
$\varepsilon = \left({\begin{array}{cc} 0 & -1 \\ 1 &  0\end{array}}\right)$,
while the  $SU(3)$ indices are  suppressed.
At the limit where $\lambda_{ijk}, \lambda^{'}_{ijk},  \lambda^{''}_{ijk}$
and $\mu_{i}$  all vanish, one recovers the expression
for the R-parity preserving MSSM, with $\hat{L}_{0}$ identified as $\hat{H}_d$. 
The latter, however, should be treated as a fourth flavor
of $\hat{L}$-type, as in our notation.

In the single-VEV parametrization\cite{I},  flavor bases are chosen
such that: 1/ among $\hat{L}_\alpha$'s, only  the $\hat{L}_0$, bears a VEV;
2/  $h^{e}_{ik} (\equiv 2\lambda_{i0k} =-2\lambda_{0ik}) 
=\frac{\sqrt{2}}{v_{\scriptscriptstyle d}}{\rm diag}
\{m_{\scriptscriptstyle 1},
m_{\scriptscriptstyle 2},m_{\scriptscriptstyle 3}\}$;
3/ $h^{d}_{ik} (\equiv \lambda^{'}_{i0k}) 
= \frac{\sqrt{2}}{v_{\scriptscriptstyle d}}{\rm diag}\{m_d,m_s,m_b\}$; 
4/ $h^{u}_{ik}=\frac{-\sqrt{2}}{v_u}V_{\scriptscriptstyle 
\!C\!K\!M}^{\dag} {\rm diag}\{m_u,m_c,m_t\}$. 
The number of parameters used is minimal, and the (tree-level) mass 
matrices for {\it all} the fermions
{\it do not} involve any trilinear RPV coupling. The parametrization
hence provides a tractable framework for the analysis of the {\it complete}
thoery of SUSY with R-parity {\it without any assumption}, verses 
models of various forms of R-parity violation, as discussed in 
Ref.\cite{I}. We want to emphasize here that, unlike most of R-parity
violation studies, the approach makes {\it no assumption} 
on any RPV coupling including those from soft SUSY breaking,
and all the parameters used are uniquely definite as a set of flavor bases
is explicitly chosen. 

Here we write the neutral fermion (neutralino-neutrino) mass matrix as
\begin{equation} \label{mn}
{\cal{M}_{\scriptscriptstyle N}} 	
=  \left(
{\begin{array}{ccccccc}
{{M}_{\scriptscriptstyle 1}} & 0 &  \frac {{g}_{1}{v}_{u}}{2}
 &  -\frac{{g}_{1}{v}_{d}}{2} & 0 & 0 & 0 \\
0 & {{M}_{\scriptscriptstyle 2}} &  -\frac{{g}_{2}{v}_u}{2} & 
\frac{{g}_{2}{v}_{d}}{2} & 0 & 0 & 0 \\
 \frac {{g}_{1}{v}_{u}}{2} &   -\frac{{g}_{2}{v}_u}{2} & 0 & 
 - {{\mu}_{\scriptscriptstyle 0}} &  - {{ \mu}_{\scriptscriptstyle 1}} 
 &  - {{ \mu}_{\scriptscriptstyle 2}} &  - {{ \mu}_{\scriptscriptstyle 3}} \\
 -\frac{{g}_{1}{v}_{d}}{2} & \frac{g_{2}{v}_d}{2}
 &  - {{ \mu}_{0}} & W & 0 & Y & Z \\ 
0 & 0 &  - {{ \mu}_{\scriptscriptstyle 1}} & 0 & 0 & 0 & 0 \\
0 & 0 &  - {{ \mu}_{\scriptscriptstyle 2}} & Y & 0 & A & C \\
0 & 0 &  - {{ \mu}_{\scriptscriptstyle 3}} & Z & 0 & C & B
\end{array}}
 \right)  \; ,
\end{equation}
with the basis vector
$
\Psi^{\scriptscriptstyle T}_0=(-i\lambda',\,\, 
-i\lambda_{\scriptscriptstyle 3},\,\, 
\psi^{\scriptscriptstyle 2}_{\!\scriptscriptstyle H_u},\,\, 
\psi^{\scriptscriptstyle 1}_{\!\scriptscriptstyle L_0},\,\, ,
\psi^{\scriptscriptstyle 1}_{\!\scriptscriptstyle L_1},\,\, 
\psi^{\scriptscriptstyle 1}_{\!\scriptscriptstyle L_2},\,\, 
\psi^{\scriptscriptstyle 1}_{\!\scriptscriptstyle L_3})
$
where $-i\lambda'$ and $-i\lambda_{\scriptscriptstyle 3}$ are the
unmixed bino and neutral wino states, the rest are the neutral components
of the doublets (the upper index is a $SU(2)$ one). Parameters $A,\; B,$
and $C$, and $W,\; Y,$ and $Z$ are two groups of relevant 1-loop
contributions to be discussed below. When these parameters are set to
zero, one recover the tree-level result. In the limit where the $\mu_i$'s also 
vanish, we have the R-parity conservating result and 
$\psi^{\scriptscriptstyle 1}_{\!\scriptscriptstyle L_i}$'s 
are exactly the (three) neutrino states, 
$\psi^{\scriptscriptstyle 2}_{\!\scriptscriptstyle H_u}$ and
$\psi^{\scriptscriptstyle 1}_{\scriptscriptstyle L_0}$ are the 
two higgsino states. Having $\mu_i$'s small compare to the electroweak
scale alos implies the 
$\psi^{\scriptscriptstyle 2}_{\!\scriptscriptstyle L_i}$'s are
basically the physical charged leptons
 ($\ell_i = e$, $\mu$, and $\tau$)\cite{I}. Note that the tree-level result
here is {\it exact}. We have made no assumption on the other VEV's, 
for example. We simply go to a base where they vanish. As for the 1-loop
contributions, we neglect those related to the first family --- the ``$\nu_e$".
The latter is expected to be not relevant for the Super-K atmospheric
neutrino result. 

At tree level, only two eigenvalues of ${\cal{M}_{\scriptscriptstyle N}}$
remain zero. One neutrino state acquires a mass through a ``see-saw"
mechanism.  The (massive) neutrino state is given by
\begin{equation} \label{nu5}
\left|\nu_{\scriptscriptstyle 5}\right\rangle = 
\frac{\mu_{\scriptscriptstyle 1}}{\mu_{\scriptscriptstyle 5}}
\left|\psi^{\scriptscriptstyle 1}_{\scriptscriptstyle L_1}\right\rangle
+ \frac{\mu_{\scriptscriptstyle 2}}{\mu_{\scriptscriptstyle 5}}
\left|\psi^{\scriptscriptstyle 1}_{\scriptscriptstyle L_2}\right\rangle
+  \frac{\mu_{\scriptscriptstyle 3}}{\mu_{\scriptscriptstyle 5}}
\left|\psi^{\scriptscriptstyle 1}_{\scriptscriptstyle L_3}\right\rangle \; ,
\end{equation}
where 
\begin{equation}
\mu_{\scriptscriptstyle 5} = \sqrt{\mu_{\scriptscriptstyle 1}^2 
+\mu_{\scriptscriptstyle 2}^2 +\mu_{\scriptscriptstyle 3}^2} \; .
\end{equation}
 The important point to note here is that 
(at tree-level) the massive neutrino is an admixture of the three
$\psi^{\scriptscriptstyle 1}_{\scriptscriptstyle L_i}$'s,
which correspond to $\nu_e$, $\nu_\mu$, and $\nu_\tau$ at the limit of
small $\mu_i$'s; and the $\nu_\mu$ and $\nu_\tau$ content, for example,
of the massive neutrino is given by 
$\frac{\mu_{\scriptscriptstyle 2}}{\mu_{\scriptscriptstyle 5}}$ and
$\frac{\mu_{\scriptscriptstyle 3}}{\mu_{\scriptscriptstyle 5}}$ respectively.
Actually, one can extract the effective neutrino mass matrix by considering
${\cal{M}_{\scriptscriptstyle N}}$ of Eq.(\ref{mn}) in the $3+4$ block form
$\left(
\begin{array}{cc}
\cal{M} & \xi^{\scriptscriptstyle T} \\
\xi & m_\nu^{\scriptscriptstyle 0}
\end{array}
 \right)$,  
with the explicit ``see-saw"  type structure. Up to first order, we have
\begin{equation} \label{ss}
m_\nu = - \xi {\cal{M}}  \xi^{\scriptscriptstyle T} + 
m_\nu^{\scriptscriptstyle 0} \; .
\end{equation}
Since the ``heavy" neutrino should not contain much of $\nu_e$, we assume 
${\mu_{\scriptscriptstyle 1}}$ to be negligible. To simplify the analysis,
we drop  any reference to the $\nu_e$ state and contract the mass
matrices accordingly; {\it i.e.} $m_\nu^{\scriptscriptstyle 0}$ and 
$\xi$ are now considered as $2\times 2$ and $2\times 4$ matrices. When only 
tree-level contribution is considered, we have $m_\nu^{\scriptscriptstyle 0}=0$. 
Then $m_\nu$ is given from the pure ``see-saw" contribution and
the result is a matrix of the form
\begin{equation} \label{ab}
\Xi =
\left( \begin{array}{cc}
a^2 & ab \\
ab & b^2
\end{array}\right) \; .
\end{equation}
This form of matrix is particularly important in our discussions below. It
suffices to note that it is diagonalized by a rotation of angle $\theta$ with
$\tan\!\theta = a/b$, to give eigenvalues $0$ and $a^2+b^2$. 
In the case here, $a:b=\mu_{\scriptscriptstyle 2}:\mu_{\scriptscriptstyle 3}$,
and the nonvanishing eigenvalue is the neutrino mass
\begin{equation} \label{mtree}
m_{\nu_{\scriptscriptstyle 5}} \simeq  -  \frac {1}{2}
\frac{ {\mu}_{5}^{2} {v}^{2} \cos^2\!\!\beta 
\left( x{g}_{\scriptscriptstyle 2}^{2} 
+ {g}_{\scriptscriptstyle 1}^{2} \right) }
{\mu_{\scriptscriptstyle 0} \left[ 2xM_{\scriptscriptstyle 2}
 \mu_{\scriptscriptstyle 0} -
 \left( x{g}_{\scriptscriptstyle 2}^{2}+{g}_{1\scriptscriptstyle }^{2}\right) 
{v}^2 \sin\!{\beta}\cos\!{\beta} \right] }\; ,  
\end{equation}
where we have substituted $v_{\scriptscriptstyle d}=v\cos\!{\beta}$, 
$v_{\scriptscriptstyle u}=v\sin\!{\beta}$,
and $M_{\scriptscriptstyle 1}=xM_{\scriptscriptstyle 2}$.
If this is the dominating contribution to $m_\nu$, for the neutrino 
oscillation scenario we are interested in here,  we expect
$m_{\nu_{\scriptscriptstyle 5}}^2 \simeq \Delta\!m^2_{\rm atm}$.
Eq.(\ref{mtree}) then gives 
\begin{equation} \label{mub}
{\mu_{\scriptscriptstyle 5}}\cos\!{\beta} \sim 10^{-4}\, \mbox{GeV}\; .
\end{equation}
Furthermore, we have
\begin{equation} \label{23}
\sin\!{2\theta}_{\rm atm} = \frac{2{\mu_{\scriptscriptstyle 2}}
{\mu_{\scriptscriptstyle 3}}} {\mu_{\scriptscriptstyle 5}^2} \; ,
\end{equation}
giving
\begin{equation} \label{63}
\frac{{\mu_{\scriptscriptstyle 2}}} {{\mu_{\scriptscriptstyle 3}}} 
\mathrel{\raise.3ex\hbox{$>$\kern-.75em\lower1ex\hbox{$\sim$}}}  0.6358 \; .
\end{equation}

From the above analysis, we have  arrived at the first interesting point 
about the flavor structure : fitting the neutrino oscillations suggests  
\[
{\mu_{\scriptscriptstyle 1}} \ll {\mu_{\scriptscriptstyle 2}} \sim 
{\mu_{\scriptscriptstyle 3}}\;\; (\,\ll {\mu_{\scriptscriptstyle 0}}\, ) \; .
\]
Remember that our $\mu_\alpha$'s as mass-couplings for the 
$\hat{L}_{\alpha} \hat{H}_u$ terms are written in the base where the three 
$\hat{L}_i$'s correspond, up to negligibly small perturbation by the 
$\mu_i$'s, to the $\ell_i$ mass eigenstates\cite{I}.
The flavor structure among the $\mu_i$'s then marks a clear 
deviation from the simple hierarchical pattern down the three families.

At 1-loop level, all three neutrinos are expected to be massive.
Studies of the 1-loop contributions to neutrino masses from R-parity 
violation has been quite a popular topic\cite{nurr,nuR}. However, they are 
usually done in the context of models of specific forms of R-parity 
violation, with typically assumptions on the vanishing of the bilinear
RPV terms and/or ``sneutrino " VEV's. Otherwise, they are studied as
contributions to individual entries of the neutral fermion mass matrix
with all family indices taken to be in the mass eigenstate basis of the
charged lepton, a feature that makes them incompatible with the analysis
of the tree-level contributions from the bilinear RPV terms as the 
latter modify the nature of the charged lepton states too. A consistent 
framework to handle both the tree and 1-loop contributions together 
without any assumption on the nature of R-parity violation is provided
by the single-VEV parametrization summaried above. Reader are referred
to ref.\cite{I} for more details. In the limit of small $\mu_i$'s, as is the
case for the neutrino oscillation scenario addressed here. The single-VEV
basis for the neutral fermion mass matrix $\cal{M}_N$ is an excellent
approximation of the $\nu_e$-$\nu_\mu$-$\nu_\tau$ basis.  Hence, we
can simply adopt the some of the available results for expressions
of mass matrix entries like $A$, $B$, and $C$ in Eq.(2) in terms of the 
trilinear RPV couplings. Entries like $W$, $Y$, and $Z$ are commonly
neglected. Their stucture are similar to the other ones. We will check
and show that they are really negligible, at least in the present context.

The $\lambda$ and $\lambda^{'}$ RPV couplings give rise to 
the following
1-loop contributions to the $4\times 4$ Majorana mass matrix of the
$\psi^{\scriptscriptstyle 1}_{\scriptscriptstyle L_\alpha}$'s
\begin{eqnarray}
(m^{\scriptscriptstyle LL})^\ell_{\alpha \beta} &=& \frac{1}{16\pi^2} 
\lambda_{i \alpha j} \lambda_{j \beta i} \left(
\frac{A^{\ell}_{j}m^\ell_{i}}{\tilde{m}^2_{\ell_j}} + 
\frac{A^{\ell}_{i}m^\ell_{j}}{\tilde{m}^2_{\ell_i}} \right) \; , \\ 
(m^{\scriptscriptstyle LL})^{q}_{\alpha \beta} &=& \frac{3}{16\pi^2} 
\lambda^{'}_{i\alpha j} \lambda^{'}_{j \beta i} \left(
\frac{A^d_{j}m^d_{i}}{\tilde{m}^2_{q_j}} + 
 \frac{A^d_{i}m^d_{j}}{\tilde{m}^2_{q_i}} \right) \; .
\end{eqnarray}
Here, $m^\ell_{i}$ and $m^d_{i}$ are the charged lepton and down-sector
quark masses respectively, $A^{\ell}_{i}$ and $A^d_{i}$  their corresponding
slepton and squark mixing terms, and  ${\tilde{m}^2_{\ell_i}}$ and
${\tilde{m}^2_{q_i}}$ the corresponding slepton and squark loop mass factors. 
These 1-loop results are written down for  couplings in
the mass eigenstate bases of the quarks and charged leptons. Recall that
in the small $\mu_i$ limit of interest here, the bases coincide with that of
the single-VEV parametrization. We have dropped contributions involving 
off-diagonal $A^d$ or $A^\ell$, adopting the common assumption that 
they are negligible. Matrix $m^{\scriptscriptstyle LL}$ 
corresponds to the lower $4\times 4$-block of 
${\cal{M}_{\scriptscriptstyle N}}$ given in Eq.(\ref{mn}).

Ref.\cite{CKKL} has concentrated on the effects of the 
$\lambda^{'}_{3i3}$ couplings. It is obvious that they give the 
dominating contributions to $(m^{\scriptscriptstyle LL})^{q}$, 
for reasonable values of the SUSY parameters.
Amazing enough, their contribution to $m_\nu^{\scriptscriptstyle 0}$
is also of the form given by $\Xi$ [{\it cf.} Eq.(\ref{ab})]. Explicitly, 
we have
\begin{equation} \label{mnu1q}
(m_\nu^{\scriptscriptstyle 0})^q \simeq \frac{3}{8\pi^2}
\frac{m^2_b}{M_{\scriptscriptstyle S\!U\!S\!Y}}
\left( \begin{array}{cc}
 \lambda^{'2}_{\scriptscriptstyle 3\!2\!3} & 
\lambda^{'}_{\scriptscriptstyle 3\!2\!3}
 \lambda^{'}_{\scriptscriptstyle 3\!3\!3} \\
\lambda^{'}_{\scriptscriptstyle 3\!2\!3} 
\lambda^{'}_{\scriptscriptstyle 3\!3\!3} & 
 \lambda^{'2}_{\scriptscriptstyle 3\!3\!3}
\end{array}\right) \; ,
\end{equation}
where $M_{\scriptscriptstyle S\!U\!S\!Y}$ denotes some SUSY scale
mass. The $ \lambda^{'}$'s at the order of $10^{-5}$ is then relevant 
for our neutrino oscillation scenario. At $10^{-4}$, they could be the 
dominating contribution to $m_\nu$. In that case, fitting
$\sin\!{2\theta}_{\rm atm}$ requires 
$\frac{\lambda^{'}_{\scriptscriptstyle 3\!2\!3}}
{ \lambda^{'}_{\scriptscriptstyle 3\!3\!3}} \mathrel{\raise.3ex\hbox{$>$\kern-.75em\lower1ex\hbox{$\sim$}}}  0.6358$
[{\it cf.} Eqs.(\ref{23}) and (\ref{63})].

The contributions from the lepton loops are actually more interesting.
Because of the antisymmetry of the first two indices in the $\lambda$'s,
we have 
\begin{equation}
(m_\nu^{\scriptscriptstyle 0})^\ell \simeq \frac{1}{8\pi^2
M_{\scriptscriptstyle S\!U\!S\!Y}}
\left( \begin{array}{cc}
m_\tau^2 \lambda_{\scriptscriptstyle 3\!2\!3}^2 &
- m_\mu m_\tau \lambda_{\scriptscriptstyle 3\!2\!2}
 \lambda_{\scriptscriptstyle 3\!2\!3}  \\
- m_\mu m_\tau \lambda_{\scriptscriptstyle 3\!2\!2}
 \lambda_{\scriptscriptstyle 3\!2\!3}
&  m_\mu^2 \lambda_{\scriptscriptstyle 3\!2\!2}^2
\end{array}\right) \; .
\end{equation}
It still has the same matrix form of $\Xi$ but now the $a:b$ ratio
is expected to be at least $m_\tau : m_\mu$, meaning that fitting
$\sin\!{2\theta}_{\rm atm}$ using this as the dominating contribution
is hopeless. Having $\lambda$'s of the order $10^{-4}$ would still make
them relevant to the neutrino oscillation scenario, generating the lighter
neutrino masses. It is of particular interest to note the strong anti-hierarchy
among the matrix elements. The antisymmetry of the $\lambda$'s uniquely
forbids the $m_\tau$ contributions to the $\nu_\tau$ mass terms, 
singling out the contribution to the $\nu_\mu$ mass.

So far, we discuss only the part of $m^{\scriptscriptstyle LL}$
contributing to $m_\nu^{\scriptscriptstyle 0}$, without $\nu_e$. 
They correspond to 
parameters $A$, $B$, and $C$ in Eq.(\ref{mn}). The other parts of 
$m^{\scriptscriptstyle LL}$ actually have larger contributions, but to
parameters $W$, $Y$, and $Z$. For example, 
 $(m^{\scriptscriptstyle LL})^q_{00} \simeq  \frac{3}{8\pi^2}
\frac{m^2_b}{M_{\scriptscriptstyle S\!U\!S\!Y}}
  \lambda^{'2}_{\scriptscriptstyle 3\!0\!3} \sim 10^{-3} \, \mbox{GeV}$ 
(note: $\lambda^{'}_{\scriptscriptstyle 3\!0\!3}$
 {\it is} the bottom Yukawa) contributes to $W$,
which then marks a negligible contribution to $\cal{M}$ in Eq.(\ref{ss}).
Similarly,  $(m^{\scriptscriptstyle LL})^q_{0i} \simeq  \frac{3}{8\pi^2}
\frac{m^2_b}{M_{\scriptscriptstyle S\!U\!S\!Y}}
\lambda^{'}_{\scriptscriptstyle 3\!0\!3}  
 \lambda^{'}_{\scriptscriptstyle 3\!i\!3}$ 
contributes to $Y$ and $Z$. For
$\lambda^{'}_{\scriptscriptstyle 3\!i\!3}
\sim 10^{-4}$, $Y$ and $Z \sim 10^{-7} \, \mbox{GeV}$. 
However, they come into $m_\nu$ through $\xi$ in  Eq.(\ref{ss}), like
the $\mu_i$'s. Comparison with Eq.(\ref{mub}) then indicates that
they are really negligible. Similar contributions from 
$(m^{\scriptscriptstyle LL})^\ell$ have even further suppressions from
antisymmetry of the $\lambda$'s indices. All loop contributions to
mass terms involving $\nu_e$ are expected to be much suppressed as a
result of the flavor structure discussed below.

After analyzing the various sources of neutrino masses, let us look at 
the flavor stucture more carefully. The idea of the approximate 
flavor symmetry approach\cite{H} is to 
attach a suppression factor to each chiral multiplet, 
such as $\varepsilon_{\!\scriptscriptstyle L_i}$'s to the $L_i$'s and 
$\varepsilon_{\!\scriptscriptstyle E^c_i}$'s to the 
$E^{\scriptscriptstyle C}_i$'s;
here multiplets are {\it not} mass eigenstates. The hierarchical structure of
the Yukawa couplings giving rise to the fermion masses is then described
by the hierarchy among the $\varepsilon$'s for the multiplets of the same
type. We want to see if fitting the neutrino oscillation scenario could still 
be compatible with the overall flavor structure, and if we could learn
something more about the scenario, using the approach.

With SUSY, the suppression factors go with the superfields.
Without R-parity, we have four  $ \hat{L}$ flavors with
$\varepsilon_{\!\scriptscriptstyle L_0} \gg 
\varepsilon_{\!\scriptscriptstyle L_i}$.
Let us start by considering, for example, two leptonic doublets
$L^{'}_i$ and $L^{'}_j$
going with flavor factors $\varepsilon_{\!\scriptscriptstyle L_i}$ and
$\varepsilon_{\!\scriptscriptstyle L_j}$. The corresponding 
$m_\ell m_\ell^{\dag}$ matrix again has the form $\Xi$ of Eq.(\ref{ab})
with $a:b = \varepsilon_{\!\scriptscriptstyle L_i} : 
\varepsilon_{\!\scriptscriptstyle L_j}$, to be diagonalized by a 
rotation of angle given by $\tan\!\theta_{ij}= 
\varepsilon_{\!\scriptscriptstyle L_i} /
\varepsilon_{\!\scriptscriptstyle L_j}$. When  
$\varepsilon_{\!\scriptscriptstyle L_i} \ll
\varepsilon_{\!\scriptscriptstyle L_j}$, replacing $L^{'}_i$ and $L^{'}_j$
by the mass eigenstates $L_i$ and $L_j$ simply makes the
$\varepsilon_{\!\scriptscriptstyle L_i}$ and
$\varepsilon_{\!\scriptscriptstyle L_j}$ factors go along with 
$L_i$ and $L_j$ (see also Ref.\cite{H}). With 
$\varepsilon_{\!\scriptscriptstyle L_i} \sim
\varepsilon_{\!\scriptscriptstyle L_j}$, however, the factors that go 
with $L_i$ and $L_j$ would be $\sin\!2\theta_{ij} 
\sqrt{\varepsilon_{\!\scriptscriptstyle L_i}^2 + 
\varepsilon_{\!\scriptscriptstyle L_j}^2}$ and $\cos\!2\theta_{ij} 
\sqrt{\varepsilon_{\!\scriptscriptstyle L_i}^2 + 
\varepsilon_{\!\scriptscriptstyle L_j}^2}$, which are of course
still of the same order of magnitude as 
$\varepsilon_{\!\scriptscriptstyle L_i}$ and 
$\varepsilon_{\!\scriptscriptstyle L_j}$.

Our neutrino oscillation scenario, together with the known charged
lepton masses, suggests
\begin{eqnarray*}
\varepsilon_{\!\scriptscriptstyle L_1} \ll
\varepsilon_{\!\scriptscriptstyle L_2} \sim
\varepsilon_{\!\scriptscriptstyle L_3} \ll
\varepsilon_{\!\scriptscriptstyle L_0} \; , \\
\noalign{\hbox{and}}
\varepsilon_{\!\scriptscriptstyle E^c_1} \ll
\varepsilon_{\!\scriptscriptstyle E^c_2} \ll
\varepsilon_{\!\scriptscriptstyle E^c_3}  \; .
\end{eqnarray*}
If we naively take $\frac{m_\tau}{m_t}$ as roughly the suppression factor 
for $L_3$, namely $\cos\!2\theta_{\scriptscriptstyle 23} 
\sqrt{\varepsilon_{\!\scriptscriptstyle L_2}^2 + 
\varepsilon_{\!\scriptscriptstyle L_3}^2}$, we obtain the kind
of maximum suppression in the $\lambda$ and $\lambda^{'}$ 
couplings from the approximate flavor symmetry perspective. 
This requires $\varepsilon_{\!\scriptscriptstyle E^c_3}\sim 1$, 
and hence $\varepsilon_{\!\scriptscriptstyle E^c_2}\sim 
\frac{m_\mu}{m_\tau}$.
We have then the natural suppressions:
$\lambda^{'}_{\scriptscriptstyle 3\!3\!3} \sim 
\lambda^{'}_{\scriptscriptstyle 3\!2\!3} \sim 5\times 10^{-4} $
(taking also $m_b/m_t$ into $\varepsilon_{\!\scriptscriptstyle Q_3}
\varepsilon_{\!\scriptscriptstyle D^c_3}$); $\lambda_{323} \sim 10^{-4} $;
and $\lambda_{322} \sim 10^{-5} $. From the discussions above, we 
see that only the $\lambda^{'}$'s are marginally too large. But then we have 
used a very conservative value of $100\, \mbox{GeV}$
for $M_{\scriptscriptstyle S\!U\!S\!Y}$ of Eq.(\ref{mnu1q}) above,
and some numerical factors close to unity actually go with the
$\varepsilon$'s to give the exact couplings.
It seems then the interplay among the relevant SUSY parameters could
take care of the problem. Much stronger suppression for any
coupling involving $L_1$ is obvious, hence justifying their being neglected
in our analysis here.

Now we turn to the $\mu_i$'s, though trying to understand the suppressions
of these bilinear couplings from the same framework represents a
much more ambitious program. The origin of the general $\mu$-terms may
actually related to SUSY breaking (see, for example, Ref.\cite{NP}). 
However, it worths taking
a look and see how far one can go. The flavor suppression factors 
involved in the $\mu_{\scriptscriptstyle 2}$ and $\mu_{\scriptscriptstyle 3}$
terms are  $\sin\!2\theta_{\scriptscriptstyle 23} 
\sqrt{\varepsilon_{\!\scriptscriptstyle L_2}^2 + 
\varepsilon_{\!\scriptscriptstyle L_3}^2}$ and
$\cos\!2\theta_{\scriptscriptstyle 23} 
\sqrt{\varepsilon_{\!\scriptscriptstyle L_2}^2 + 
\varepsilon_{\!\scriptscriptstyle L_3}^2}$ respectively.
 Eq.(\ref{mub}) requires
$\mu_i/ \mu_{\scriptscriptstyle 0} < 10^{-6}$, far beyond the 
suppressions of the order $\frac{m_\tau}{m_t}$.
Nevertheless, we have not
taken into the consideration of the rotation from the arbitary $L_\alpha^{'}$'s 
one attach the $\varepsilon_{\!\scriptscriptstyle L_\alpha}$'s 
to our single-VEV bases.  A previous study\cite{B}
has  shown that an alignment between the $\mu_\alpha$ and and the
relevant soft SUSY breaking terms results in a similar alignment of
the VEV's among the $\hat{L}_\alpha$'s. Translated into the single-VEV
parametrization, that implies a suppression of the $\mu_i$'s. An
approximate  flavor symmetry would provide the kind of alignment,
though a detailed study is certainly called for to establish that a
good enough alignment could be obtained for our purpose. 
Assuming that is the $\mu_i$'s being the dominating
contribution to $m_\nu$ with the right order to fit the oscillation scenario 
concerned here, then we have, from Eqs.(\ref{23}) and 
(\ref{63}) and the above discussions,
\begin{equation}
\sin\!{2\theta}_{\rm atm}
\simeq 2 \sin\!2\theta_{\scriptscriptstyle 23}  
\cos\!2\theta_{\scriptscriptstyle 23} 
\simeq \frac{ 4 \varepsilon_{\!\scriptscriptstyle L_2}
\varepsilon_{\!\scriptscriptstyle L_3}
 ( \varepsilon_{\!\scriptscriptstyle L_3}^2
- \varepsilon_{\!\scriptscriptstyle L_2}^2 ) }
{ ( \varepsilon_{\!\scriptscriptstyle L_2}^2 
+ \varepsilon_{\!\scriptscriptstyle L_3}^2 )^{2} }\; .
\end{equation}
A quite interesting result.

Neutrino phenomenology could be more complicated than the simple
scenario considered here. In such situations, the RPV couplings discussed
above may not be as restricted. In particular, the $\mu_i$'s, especially 
$\mu_{\scriptscriptstyle 3}$, may be substantial and their contribution
to $m^{\ell}_i$'s not negligible.  Unlike the case here, the  larger 
values of the admissible RPV couplings might easily fit in the approximate 
flavor symmetry perspective. However, an analysis of the flavor structure issue
there would be inevitably more involved. We hope to report on a more complete 
analysis in a future publication.

In summary, from our brief analysis here, we have illustrated a few
interesting issues in the flavor structure of SUSY without R-parity.
Though the suppressions of the RPV couplings for fitting the
limiting scenario of neutrino oscillations, motivated by the recent
Super-K result, are much stronger than other phenomenological 
constraints\cite{rpv,BFK}, there is still room for fitting them into
an approximate flavor symmetry perspective. Success of the latter
is a strong indication that the R-parity (or lepton number) violating
couplings  are ``naturally" small, as the light fermion masses are,
and their explanation most probably lies under a common theory of
flavor structure.
 
M. Bisset and X.-G. He are to be thanked for discussions.
This work was supported in part by the U.S. Department of Energy,
under grant DE-FG02-91ER40685.

\end{document}